\documentstyle[12pt]{article}

\begin{document}
\def\la{\mathrel{\mathpalette\fun <}}
\def\ga{\mathrel{\mathpalette\fun >}}
\def\fun#1#2{\lower3.6pt\vbox{\baselineskip0pt\lineskip.9pt
        \ialign{$\mathsurround=0pt#1\hfill##\hfil$\crcr#2\crcr\sim\crcr}}}
\newcommand {\eegg}{e^+e^-\gamma\gamma~+\not \! \!{E}_T}
\newcommand {\mumugg}{\mu^+\mu^-\gamma\gamma~+\not \! \!{E}_T}
\renewcommand{\thefootnote}{\fnsymbol{footnote}}
\bibliographystyle{unsrt}
\vspace{2mm}

\begin{center}

{\Large \bf A New Messenger Sector for Gauge Mediated Supersymmetry Breaking}
\vspace{1cm}

{\bf R. N. Mohapatra$^{{\rm a)}}$ and S. Nandi$^{{\rm b)}}$}
 \vspace{0.3cm}

{\em a) Department of Physics, University of Maryland, College Park, Md-20742\\

b) Department of Physics, Oklahoma State University, Stillwater, OK-74078 }

\end{center}

\vspace{2mm}
\begin{abstract}
                                                                    
We propose a new class of gauge mediated supersymmetry breaking (GMSB) models 
where the standard model gauge group is embedded into the gauge group
$SU(2)_L\times U(1)_{I_{3R}}\times U(1)_{B-L}$ (or $SU(2)_L\times
SU(2)_R\times U(1)_{B-L}$) at the supersymmetry breaking scale $\Lambda_S$.
The messenger sector in these models can consist of color singlet but
$U(1)_{I_{3R}}$ and $U(1)_{B-L}$ non-singlet fields. The distinguishing 
features of such models are: (i) exact R-parity conservation, 
(ii) non-vanishing neutrino masses and (iii) a solution to the SUSYCP problem. 
We present a complete hidden plus messenger
sector potential that leads to the desired supersymmetry breaking
pattern. The simplest version of these models predicts the existence
of very light gluinos which can be made heavier by a simple modification
of the supersymmetry breaking sector.

\end{abstract}
 
\renewcommand{\thefootnote}{\arabic{footnote})}
\setcounter{footnote}{0}
\addtocounter{page}{-1}
\baselineskip=24pt

Understanding the origin and nature of supersymmetry breaking is one of 
the major research areas in particle physics right now. The general
strategy is to postulate the existence of a hidden sector where
 supersymmetry is assumed to be broken and then have it 
transmitted to the visible sector. A currently popular way of
transmitting the supersymmetry breaking is to use the standard model 
gauge interactions in which case one can choose the SUSY breaking scale
to be $\simeq 10-100$ TeV or so. These models are known as gauge mediated
supersymmetry breaking (GMSB) models\cite{dine}. They are attractive 
for two primary reasons: one is that they 
lead to degenerate squark and slepton masses at the SUSY breaking 
scale $\Lambda_S$ which then provides a natural solution to the flavor 
changing neutral current problem of the low energy supersymmetry models. 
Secondly, these models are extremely predictive so that one can have a genuine
hope that they can be experimentally tested\cite{GMSB} 
in the not too distant future.

There are however several drawbacks in the usual construction of 
these models: first, one uses extra vectorlike quarks and 
leptons whose sole purpose is to transmit
the supersymmetry breaking from the hidden to the visible sector; secondly,
it has been argued\cite{randall} that in explicit models of supersymmetry
breaking in the hidden sector, the lowest vacuum breaks color and finally  
the lightest of the messenger fields is a heavy stable particle which
may lead to cosmological difficulties. Finally, these class of models do
not address some other generic problems of the MSSM such as the existence
of R-parity breaking interactions that lead to 
arbitrary couplings for the unwanted baryon and lepton number 
violation. Our goal in this paper is to propose an
alternative messenger sector which eliminates some of these problems of
the GMSB models.

Our model is based on the electroweak gauge group $SU(2)_L\times U(1)_{I_{3R}}
\times U(1)_{B-L}$ and uses the color singlet but $B-L$ and $I_{3R}$ 
non-singlet vectorlike fields as the messenger fields in the simplest
version. These fields not only play
the role of messenger fields but also serve to break the above gauge group
down to that of the standard model group. Secondly, the above group
automatically guarantees the conservation of R-parity\cite{moh} so that
the model conserves baryon and lepton number automatically, a property
that made the standard model so attractive and is sadly missing in MSSM. 
It also leads to
nonzero neutrino masses via the usual see-saw mechanism. Moreover,
since the messenger fields couple to the right-handed neutrinos 
they are unstable and therefore do not cause any cosmological problems. 
Finally, since in this model, the Majorana masses for the gauginos vanish upto
one loop level as does the A-parameter, one has a solution to the
SUSYCP problem. Key predictions of
this model are : (i) the existence of very light gluinos, which, as far as we
know, are not conclusively ruled out\cite{farrar}; (ii) the existence of
a chargino lighter than the W-boson and (iii) photino as the lightest
neutralino with mass in the 1/2 to 1 GeV range. When 
photinos are the lighter than gluinos, big bang nucleosynthesis
puts an upper bound on the gravitino mass of 40 eV. We also give an explicit
model for supersymmetry breakdown in the hidden sector and its transmission
to the visible sector.

\noindent{\it The Model}

Under the gauge group $SU(2)_L\times U(1)_{I_{3R}}\times U(1)_{B-L}$,
the quarks and leptons (including the right-handed
neutrinos, $\nu^c$) transform as $Q (2, 0, 1/3)~~$; $L (2, 0, -1)~~$;
$u^c (1, -1/2, -1/3)~~$; $d^c (1, +1/2, -1/3)~~$; $e^c (1, +1/2, +1)~~$; 
$\nu^c (1, -1/2, +1)$ and the two MSSM Higgs doublets
as $H_u (2, +1/2, 0)$ and $ H_d (2, -1/2, 0)$. In addition to these, we
add the fields $\delta $ and $\bar{\delta}$  with quantum numbers
$\delta (1, +1, -2)$ and $\bar{\delta} (1, -1, +2)$ 
that will break the $U(1)_{I_{3R}}\times U(1)_{B-L}$ symmetry
down to the $U(1)_Y$ of the standard model. The interesting point 
 illustrated in this simplest version of our
model is that the same $\delta$ and $\bar{\delta}$ fields which
play a role in supersymmetry breaking also play the
role of messenger fields.

To study the general profile of the model, we start with the following
superpotential $W = W_m + W_g$ where $W_m = \lambda S\delta \bar{\delta}$
corresponds to the messenger sector and is part of the complete unified
hidden and messenger sector superpotential (described in the subsequent 
section). $W_g$ describes the visible sector of our model.
\begin{eqnarray}
W_g = h_u QH_u u^c + h_d Q H_d d^c + h_e L H_d e^c + h_{\nu} L H_u \nu^c
+ \mu H_u H_d + f \delta \nu^c\nu^c
\end{eqnarray}
We will show  that $F_S$, $S$, $\delta$ and $\bar{\delta}$ 
acquire nonzero vacuum expectation values (vev) so that supersymmetry 
as well as $U(1)_{B-L}\times U(1)_{I_{3R}}$ are broken at the same scale. 
As a result, the supersymmetry breaking scale and the $B-L$ breaking scale
get linked to each other and cannot be arbitrarily adjusted in the physics
discussion. It is also worth pointing out that 
the fields ($\delta$ and $\bar{\delta}$) are essential for the supersymmetry 
breaking. Next, we note that the same mechanisms 
that give mass to the gauginos at the one loop
and sfermions at the two loop level in the usual GMSB models help to
give mass to the $B-L$ and $I_{3R}$ gauginos ($\lambda_{B-L, I_{3R}}$)
at the one loop and all sfermions at the two loop level; 
 The mass matrix for the $I_{3R}$ and $B-L$ gaugino-
$\tilde{\delta}$, $\tilde{\bar{\delta}}$ 
is given in the basis $(\lambda_{B-L},\lambda_{R},
\tilde{\delta}, \tilde{\bar{\delta}})$ by:
\begin{eqnarray}
M =  \frac{1}{4\pi} \left( \begin{array}{cccc}
-\alpha_{B-L}\Lambda_S & \sqrt{\alpha_{B-L}\alpha_R}\Lambda_S & -\sqrt{2}4\pi
g_{B-L}v_{\delta} & \sqrt{2}4\pi g_{B-L}v_{\bar{\delta}} \\
\sqrt{\alpha_{B-L}\alpha_R}\Lambda_S & -\alpha_R\Lambda_S & \sqrt{2}4\pi
g_R v_{\delta} & -\sqrt{2} 4\pi g_R v_{\bar{\delta}} \\
-\sqrt{2}4\pi g_{B-L}v_{\delta} & \sqrt{2}4\pi g_R v_{\delta} & 0 & <S> \\
\sqrt{2} 4\pi g_{B-L} v_{\bar{\delta}} & -\sqrt{2} 4\pi g_R v_{\bar{\delta}}
& <S> & 0 \\
\end{array} \right)
\end{eqnarray}
where $\Lambda_S = <F_S>/<S>$. It is clear from this mass matrix that it
has a zero eigenstate corresponding to the $\lambda_Y\equiv (g^{-1}_{B-L}
\lambda_{B-L} + g^{-1}_{R}\lambda_{I_{3R}})$. In other words, in the
language of the MSSM, $M_1=0$.
The masses of the $\tilde{\nu^c}$ arise at
the tree level and are therefore of order $\Lambda_S$ as are the masses of
the right-handed neutrinos. Turning now to the remaining sfermions, we find
that:
\begin{eqnarray}
M^2_{\tilde{F}}\simeq 2[x^2_{F}  \left(\frac{\alpha_{B-L}}{4\pi}\right)^2 
\Lambda^2_S + y^2_{F} \left( \frac{\alpha_R}{4\pi}\right)^2\Lambda_S^2]
\end{eqnarray}
where $x_F$ and $y_F$ denote the $\frac{B-L}{2}$ and $I_{3R}$ 
values for the different
superfields $F$ (both matter as well as Higgs). It is therefore clear that
the good FCNC properties of the usual GMSB are maintained in this class
of models. Furthermore the spectrum of squarks and sleptons here is very
different from that of the usual GMSB models, where messenger fields carry
color.

There are already several differences from the usual GMSB models apparent
at this stage: first is that Majorana masses for the usual Winos and the
Binos are vanising to this order as are the masses of the gluinos. We will
address the question of their masses later. The next point to note is that
unlike in the usual GMSB models, the squark and slepton masses are of same
order (upto their B-L and $I_{3R}$ values). The renormalization group
extrapolation of $M^2_{H_u}$ from the scale $\Lambda_S$ to the weak scale makes
it negative i.e.
$M^2_{H_u}-M^2_{H_d}\simeq -\frac{3}{8\pi^2} M^2_{\tilde{t}} ln\frac{\Lambda_S}
{M_{\tilde{t}}}$.
An important new point is that since $M^2_{\tilde{t}}$ to start with is down
compared to the prediction of the usual GMSB models,
by a factor roughly $(\alpha_{B-L}/\alpha_c)^2\sim 1/36$ or so,
for $\Lambda_S\sim 100$ TeV, the value of $M^2_{H_u}$ at the weak scale
is of the order of $-(100~GeV)^2$. This makes further fine tuning 
unnecessary to achieve electroweak symmetry breaking unlike in the
usual GMSB models where the $\mu^2$ must be fine tuned to get the weak
scale right. Furthermore, our model predicts that the MSSM A-parameter
is also zero upto one-loop level.

The $B\mu$ term which leads to nonzero
Majorana masses for the Wino and Bino arises at the two loop level
from the Feynman diagram that involves the Majorana mass for the
$\lambda_{I_{3R}}$ and the $\mu$ term and
 leads at three loop level to Majorana masses for the Wino and the Bino. 
Note that electroweak symmetry
breaking also leads to Dirac type masses for the Wino. 
Coming to the gluino masses, they arise at the one loop level via the
top and stop intermediate states and can be estimated to be 
\begin{eqnarray}
M_{\tilde{G}}\simeq \frac{\alpha_s}{4\pi} \frac{m^2_t\mu cot \beta}{M^2}
\end{eqnarray}
where M is the larger of $(m_t, m_{\tilde{t}})$
Thus the gluino mass in this model is of the order of a  GeV. Perhaps
such low values of the gluino mass are not ruled out at present\cite{farrar}.
It is however worth pointing out that the light gluinos are only a
prediction of the simplest version of our model and we show below
how the superpotential for the hidden sector can be trivially modified
to make the gluinos heavy.

The fact that this model leads to see-saw model for neutrino masses and
exact R-parity conservation have already been discussed in the literature
and will not be repeated here. The values of left-handed neutrino masses
can be adjusted by ``dialing'' the values of the Yukawa coupling constants
$h_{\nu}$ in Eq. 1.

\noindent{\it An explicit model for the Hidden sector}

Let us now present an explicit model that leads to
a vev for $F_S$ and $S$ used in the previous section. This discussion is
nontrivial for the following reasons. While it is easy to costruct a
superpotential that leads to a singlet having  nonzero vevs as above, it is not 
simple to communicate the supersymmetry breaking to the visible 
sector\cite{randall}.
Furthermore, our goal is somewhat different from the usual GMSB scenario
in that we want the messenger fields to develop a vev at the supersymmetry
breaking scale. It turns out that one can construct a rather simple
superpotential that unifies the hidden sector and the messenger sector
while simultanoeusly breaking the gauge symmetry down to the standard
model.

We use three pairs of $\delta$ , $\bar{\delta}$ fields denoted by
$\delta$, $\delta'$ and $\delta''$ and the corresponding fields with
bar's (i.e. $\bar{\delta}$ etc) and two singlets $S,S'$\cite{chacko} 
and construct the following superpotential:
\begin{eqnarray}
W_{m+h}=\lambda S(\delta\bar{\delta} - M^2_0 +\delta''\bar{\delta}'')
+\lambda' S'\delta\bar{\delta} +
M_1(\delta''\bar{\delta}'+\delta'\bar{\delta}'') + M_2\delta''\bar{\delta}''
\end{eqnarray}
It is easy to see that for $M_1\gg M_0, M_2$, the ground state corresponds
to $F_S$, $F_{S'}$, $<\delta>=<\bar{\delta}>$ having nonzero vevs which
therefore break not only supersymmetry but also the $B-L$ gauge symmetry.
We also have $<\delta''>=<\bar{\delta}''>=0$ and these fields play the
role of the messengers. Furthermore, the role $<S>\neq 0$ is played by the
mass term $M_2$. It is also important to note that the D-terms all vanish
in the lowest ground state. As a result, fields such as
$\nu^c$, which couple to the $\delta$ fields for implementing the
see-saw mechanism will not acquire vevs in the lowest ground state
and thus do not effect the minimum found above.
Similarly, we have checked that the inclusion of the squark and slepton
fields into the potential also does not effect this minimum.

\noindent{\it Phenomenological Implications}
                                                                      
The fact that the Majorana masses for the gauginos vanish upto the one loop
level has important implications. The first point is that the model has
light gluinos as well as a light neutralino with masses in the range of
1/2 to 1 GeV. It has been argued\cite{farrar} that all known
data do allow a light gluino window in this range. As far
as the light neutralino is concerned, it is given by $\tilde{\gamma}
\equiv (cos \theta_W \tilde{B} + sin \theta_W \tilde{W}_3)$. As a result,
it does not couple to the Z-boson and is allowed by the
LEP Z-pole observations. Of course, which of the two (photino
($\tilde{\gamma}$) or the gluino, $\tilde{G}$) 
is lighter depends on the parameters of the model. We will therefore
comment on both cases. Since in all GMSB models the lightest
supersymmetric particle is the gravitino, the lighter of the two particles
$\tilde{\gamma}$ and $\tilde{G}$ will be the NLSP and 
will decay to the gravitino and the photon or gluon. 
Let us discuss the case of the photino being the NLSP
below: $\Gamma (\tilde{\gamma}\to \gamma +\tilde{g})\simeq 
\frac{\kappa^2 M^5_{\tilde{\gamma}}}{48\pi m^2_{\tilde{g}}}$
where $\kappa^{-1}=M_{P\ell}/\sqrt{8\pi}$. In the early universe, there
will be an abundance of photinos at some point. In order that they do not
effect the predictions of the big bang nucleosynthesis and the subsequent
evolution of the universe, its life time must be less than one second.
Using the above formula for the decay width of the photino, we conclude that,
we must have $m_{\tilde{g}}M_P\leq 5\times
10^{11}(M_{\tilde{\gamma}}/GeV)^{5/2}$ GeV$^2$. So for a one GeV NLSP, we
get $m_{\tilde{g}}\leq 40$ eV. This implies that the gravitino in this model
cannot constitute the warm dark matter of the universe.
This also leads to the conclusion that
if the photino is produced in an accelerator experiment, it
will not decay in the detector and for all practical purposes will behave
like a stable invisible particle. We can further conclude that
NNLSP (i.e. the next to the next to the lightest superparticle, denoted
here by $N_2$) whether
it be the gluino or the photino, will decay as $N_2\to q\bar{q}$+ missing 
energy due to undetected longlived NLSP.

The next particle in the mass hierarchy will be either a neutralino or
a chargino. In either case we expect the masses to be in the 60 to
100 GeV range and the particle will decay rapidly. In the neutralino case, for
arbitrary $tan \beta$, the masses are determined by the solutions of the
following cubic equation:
\begin{eqnarray}
y^3 + y (M^2_Z -\mu^2) - \mu M^2_Z sin 2\beta = 0
\end{eqnarray}
It is clear that for $tan \beta = 1$, one of the eigenvalues of this
equation is less than $M_Z$ for all values of $\mu$. However, as $tan~\beta$
gets larger, this property does not necessarily hold and depends on the
value of $\mu$. One can also show that if $\mu^2 \geq 2 M^2_Z$,
all neutralinos (except the photino) are heavier than the Z 
for arbitrary $tan \beta$. 
 
Turning to the chargino mass matrix, it is well-known that one of the
charginos in this case will be lighter than the W-boson. 
The phenomenology of this case has been discussed recently\cite{farrar}
and we simply reemphasize the fact that the ongoing LEP run can test
this prediction and at the moment, such a scenario with light gluinos
and charginos is phenomenologically viable.

Finally, we point out that it is simple to extend the above superpotential
to make the gluinos massive by including a pair of color octet (denoted
by $\theta_{1,2}$) but weak gauge singlet fields in the theory. The
superpotential in this case will be given by:
\begin{eqnarray}
W'_{m+h} = \lambda S (\delta\bar{\delta}- M^2_0+\delta''\bar{\delta}''
+\theta^2_1)+\lambda'S'\delta\bar{\delta} +M_1(\delta''\bar{\delta}'
+\bar{\delta}''\delta') +M_2\delta''\bar{\delta}''\\  \nonumber
+M_3\theta_1\theta_2 +M_4\theta^2_1
\end{eqnarray}
There will now exist one loop graphs that will make the gluino massive.
Moreover, similar method can be used to make the weak gauginos also acquire
Majorana masses\cite{chacko} by adding pairs of weak doublets.

\noindent{\it Simultaneous Breaking of Parity and Supersymmetry}

This model can be embedded into the left-right symmetric gauge group
$SU(2)_L\times SU(2)_R\times U(1)_{B-L}$ by assigning the quarks and
leptons as in the usual left-right models i.e. $Q( 2, 1, 1/3)$;
$Q^c(1, 2, -1/3)$; $L(2, 1, -1)$ and $L^c(1, 2, +1)$. The Higgs fields 
$H_{u,d}$ are embedded into two bidoublet fields $\phi_a(2, 2, 0)$ ($a=1,2$).
The fields that are chosen to break the $SU(2)_R\times U(1)_{B-L}$ symmetry
carry two units of lepton number and therefore lead to the see-saw mechanism.
They are same as in the usual supersymmetric left-right model
i.e. $\Delta (3, 1, +2)$; $\bar{\Delta} ( 3, 1, -2)$ and $\Delta^c (1, 3, -2)$;
$\bar{\Delta}^c ( 1, 3, +2)$. Note that the $\delta$ and $\bar{\delta}$
fields used earlier are part of the $\Delta^c$ and $\bar{\Delta}^c$ fields.

We will assume that both parity and supersymmetry will break at the same
scale by a mechanism similar to the one already discussed here.
In order to implement supersymmetry breaking and convey it to the visible
sector, we replace the $\delta$ fields in the Eq. 5 by the $\Delta$
and $\Delta^c$ fields in a straightforward manner and obtain 
$<\Delta^c>=<\bar{\Delta}^c>\neq 0$. The rest of the discussion is
similar to what is already given for the simple model.

 An advantage of this left-right embedding of our
model is that it provides a solution to  the strong CP problem in addition
to solving the SUSYCP as well as the R-parity problems\cite{rasin}.

There are now extra singly and doubly charged Higgs and
Higgsino fields arising from the $\Delta$ fields. The rest of 
the low energy spectrum of particles is same except that the Winos
now pick up a one loop Majorana mass from the $\Delta$ intermediate states.
As a result, the chargino need not be lighter than the W-boson.
                            
We wish to point out that since the simplest 
$SU(2)_L\times U(1)_{I_{3R}}\times U(1)_{B-L}$ model leads to
 $M_1 = M_2= M_3 = A = 0$ in the model, one has
a solution to the SUSYCP problem i.e. the electric dipole moment of the
neutron is naturally below the present experimental upper limit.
Secondly, our model does not lead to coupling constant unification
at higher scales for supersymmetry breaking scale in the 100 TeV range
unless there are additional new physics above the $\Lambda_{SUSY}$. 
                                                                    
In conclusion, we have presented a new messenger sector 
unified with the hidden sector for gauge
mediated supersymmetry breaking and explicitly constructed
a superpotential that leads to both supersymmetry breaking
as well as the breaking of extra gauge symmetries.
The model connects the supersymmetry breaking scale to the
scale of new electroweak physics such as $B-L$ or parity invariance.
The simplest version of the model predicts the existence of light gluinos
which could be used to test this version of the model\cite{farrar}. 
It is however possible to extend the model that is free of the light gluinos.
It also has the attractive features of automatic R-parity 
conservation, nonzero neutrino masses and no SUSYCP problem. 

We thank Z. Chacko and B. Dutta for discussions. The work of R. N. M.
is supported by the National Science Foundation grant no. PHY-9421386
and the work of S. N. is supported by the DOE under grant no.
DE-FG02-94-ER40852.


\end{document}